%% file: short_paper.tex
\documentclass[aps,prl,twocolumn,amsmath,amssymb,nobibnotes,superscriptaddress]{revtex4-1}
\usepackage{helvet}
\usepackage{graphicx}
\usepackage{hyperref}
\def\Tl{Tl\ensuremath{_2}Ba\ensuremath{_2}CuO\ensuremath{_{6+x}}}
\def\Tc{\ensuremath{T_\text{c}}}
\begin{document}

\title{Signatures of new $d$-wave vortex physics in overdoped \Tl\
revealed by TF-$\mu^+$SR}

\author{Jess H.\ Brewer}
\email[Correspondence to ]{jess@triumf.ca}
\author{Scott L.\ Stubbs}
\affiliation{Department of Physics and Astronomy,
 The University of British Columbia,
 Vancouver, BC, Canada V6T 1Z1}

\author{Ruixing Liang}
\author{D.\ A.\ Bonn}
\author{W.\ N.\ Hardy}
\affiliation{Department of Physics and Astronomy,
 The University of British Columbia,
 Vancouver, BC, Canada V6T 1Z1}
\affiliation{Canadian Institute for Advanced Research, 
Toronto, Ontario, Canada M5G 1Z8}

\author{J.\ E.\ Sonier}
\affiliation{Canadian Institute for Advanced Research, 
Toronto, Ontario, Canada M5G 1Z8}
\affiliation{Department of Physics, Simon Fraser University, 
Burnaby, BC, Canada V5A 1S6}

\author{W.\ Andrew MacFarlane}
\affiliation{Department of Chemistry,
 The University of British Columbia,
 Vancouver, BC, Canada V6T 1Z1}

\author{Darren C.\ Peets}
\email[Correspondence to ]{dpeets@snu.ac.kr}
\affiliation{Department of Physics and Astronomy,
 The University of British Columbia,
 Vancouver, BC, Canada V6T 1Z1}
\affiliation{Center for Correlated Electron Systems,
 Institute for Basic Science, Seoul National University,
 Seoul 151-747, Korea}

\date{\today}

\begin{abstract}
The spontaneous expulsion of applied magnetic field, the Meissner
effect, is a defining feature of superconductors; in Type-II
superconductors above the lower critical field, this screening takes
the form of a lattice of magnetic flux vortices.  Using implanted
spin-1/2 positive muons, one can measure the vortex lattice field
distribution through the spin precession and deduce key parameters of
the superconducting ground state, and thereby fundamental properties
of the superconducting pairing.  Muon spin rotation/relaxation
($\mu$SR) experiments have indeed revealed much interesting physics in
the underdoped cuprates, where superconductivity is closely related
to, or coexistent with, disordered or fluctuating magnetic and charge
excitations.  Such complications should be absent in overdoped
cuprates, which are believed to exhibit conventional Fermi liquid
behaviour.  These first transverse field (TF)-$\mu^+$SR experiments on
heavily-overdoped single crystals reveal a superfluid density
exhibiting a clear inflection point near 0.5\Tc, with a striking
doping-independent scaling.  This reflects hitherto unrecognized
physics intrinsic to $d$-wave vortices, evidently generic to the
cuprates, and may offer fundamentally new insights into their
still-mysterious superconductivity.

\end{abstract}

\maketitle


\section*{Introduction}
Charge doping of the CuO$_2$ planes tunes the occurrence of
superconductivity in the high-temperature hole-doped cuprate
superconductors between the limits of an undoped insulating
antiferromagnet and a possible conventional Fermi liquid at high
dopings.  It is appealing to try to understand how the unconventional
superconductivity evolves out of these more conventional electronic
ground states.  However, hole doping is typically effected chemically,
in the best case {\it via} the composition of a distinct,
well-separated subunit of the layered crystal structure, to leave the
planes themselves little altered structurally and the dopant site
well-shielded when a hole is promoted to the planes.  One such example
is oxygen doping in the CuO chain layer of YBa$_2$Cu$_3$O$_{7-\delta}$
(YBCO).  Unfortunately, compositional tuning is limited by
(thermodynamic) phase stability and can seldom be used to traverse the
entire superconducting phase diagram in a single system.  In this
context overdoped cuprates, those nearer the apparent Fermi liquid
regime, are rare and, moreover, relatively few have highly-ordered
CuO$_2$ planes.  For example, doping by cation substitution in
La$_x$Sr$_{2-x}$CuO$_4$ (LSCO) introduces substantial disorder
directly adjacent to the CuO$_2$ planes \cite{Eisaki2004}.  In
contrast, \Tl\ (Tl-2201) offers tunability throughout the overdoped
regime with highly-ordered, isolated, and flat CuO$_2$ planes, doped
{\it via} dilute interstitial oxygen in the distant TlO
layers~\cite{Wagner1997}, although Cu/Tl substitution in this
layer \cite{Shimakawa1993,Peets2007} may contribute an offset in
doping. The overdoping also appears to eliminate a predicted electron
Fermi surface (FS) pocket at the $\Gamma$ point, leaving only a
single, large, FS sheet \cite{Plate2005}.

Pure $d_{x^2-y^2}$-wave symmetry of the superconducting order in
Tl-2201 has been conclusively established by observation of
half-integer flux quanta at crystal boundaries in
films \cite{Tsuei1997}; line nodes are evident in
microwave \cite{Broun1997,Ozcan2006,Broun2014} and thermal
transport \cite{Hawthorn2007} measurements; and the admixture of
another pairing symmetry is unlikely because it would require
spontaneous breaking of the crystal symmetry.  However, some $\mu$SR
measurements have suggested an additionial transition at low
temperatures within the vortex state
\cite{Khasanov2007,Khasanov2008,Blasius1999}, which has been
interpreted in terms of a multiple-component order parameter.  Here,
we extend $\mu$SR studies of the vortex state of the cuprates deep
into the overdoped regime with the first transverse-field muon spin
rotation (TF-$\mu^+$SR) results on single-crystalline Tl-2201, in the
form of high-quality single crystal mosaics at a range of dopings.
Our measurements were performed at low magnetic fields, in a doping
regime free from competing charge density wave
order \cite{Ghiringhelli2012,Chang2012}, and are thus sensitive to the
intrinsic structure of $d$-wave vortices.  We show that the unusual
temperature dependence is real, and generic to the cuprates, but also
demonstrate that it is a signature of $d$-wave vortex physics rather
than a multicomponent order parameter.

\begin{figure}[htb] 
\includegraphics[width=\columnwidth,clip]{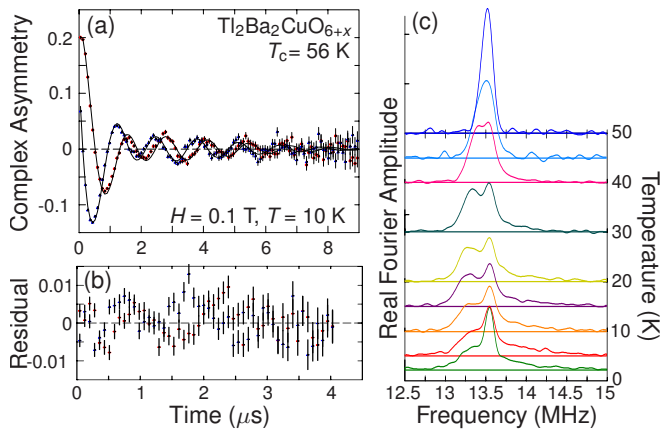} 
\caption{\label{fig:2008Tc56_1kg}Example of $\mu^+$SR
  data.  (a) Complex TF-$\mu^+$SR time spectrum (red circles: real
  part; blue triangles: imaginary part) in a rotating reference frame
  (RRF) at 0.1\,T and 10\,K on the $\Tc\approx 56$\,K Tl-2201 mosaic,
  including time-domain best fit, the residual errors of which are
  shown in (b) for the first 4\,$\mu$s where the statistics are
  highest.  (c) Fourier transforms at several temperatures.  The
  relatively sharp peak at 13.55\,MHz arises from muons stopping
  outside the sample.
}
\end{figure}

\section*{Results}

Fig.~\ref{fig:2008Tc56_1kg} shows an example of the data and time
domain fit at 10\,K on a $\Tc = 56$\,K mosaic.  Fits on all mosaics at
all fields and temperatures converged very well, and fully reproduce
the data.  Fourier transforms corresponding to the field distribution
are also shown for a selection of temperatures; the additional peak
just above the cusp is attributed to muons stopping outside the sample
and precessing about the applied field, and is accounted for in the
fits.  The high-field (high-frequency) cutoff in the lineshape is
indistinct, precluding a quantitative analysis of the in-plane
coherence length $\xi_{ab}$, but the in-plane magnetic penetration
depth $\lambda_{ab}$, which controls the linewidth, may be reliably
extracted.  Varying the fit parameters indicated that the
absolute $1/\lambda_{ab}^2$ is accurate to within $\sim 10$\%, while
its temperature dependence is robust; after a few global fits with
different choices of $\kappa_{ab} \equiv \lambda_{ab}/\xi_{ab}$ we
chose a fixed value, $\kappa_{ab} = 100$, for all remaining fits.

One strength of TF-$\mu^+SR$ in a Type-II superconductor is its
ability to determine the {\sl absolute} $\lambda$ and its inverse
square, which is proportional to the density of superconducting
carriers \cite{Sonier2000}.  Circumstances are not yet as good for
\Tl\ as for high-quality YBCO;  in particular, only small improvements
of global $\chi^2$ minimization distinguish the broadening due to
vortex lattice disorder, $\sigma_d$, which should scale with
$\lambda^{-2}(T)$, from $T$-independent broadening due to nuclear
magnetic dipoles and crystal defects, $\sigma_0$.  The amplitude $A_B$
of the background signal due to muons stopping outside the sample is
also known only from the best fit; like $\sigma_0$, it can be subtly
coupled to $\lambda^{-2}$.  These uncertainties do not alter the
temperature dependence, and have been incorporated into the quoted
$\sim 10$\% uncertainty in $1/\lambda_{ab}^2$.

\begin{figure}[htb] 
\includegraphics[width=\columnwidth,clip]{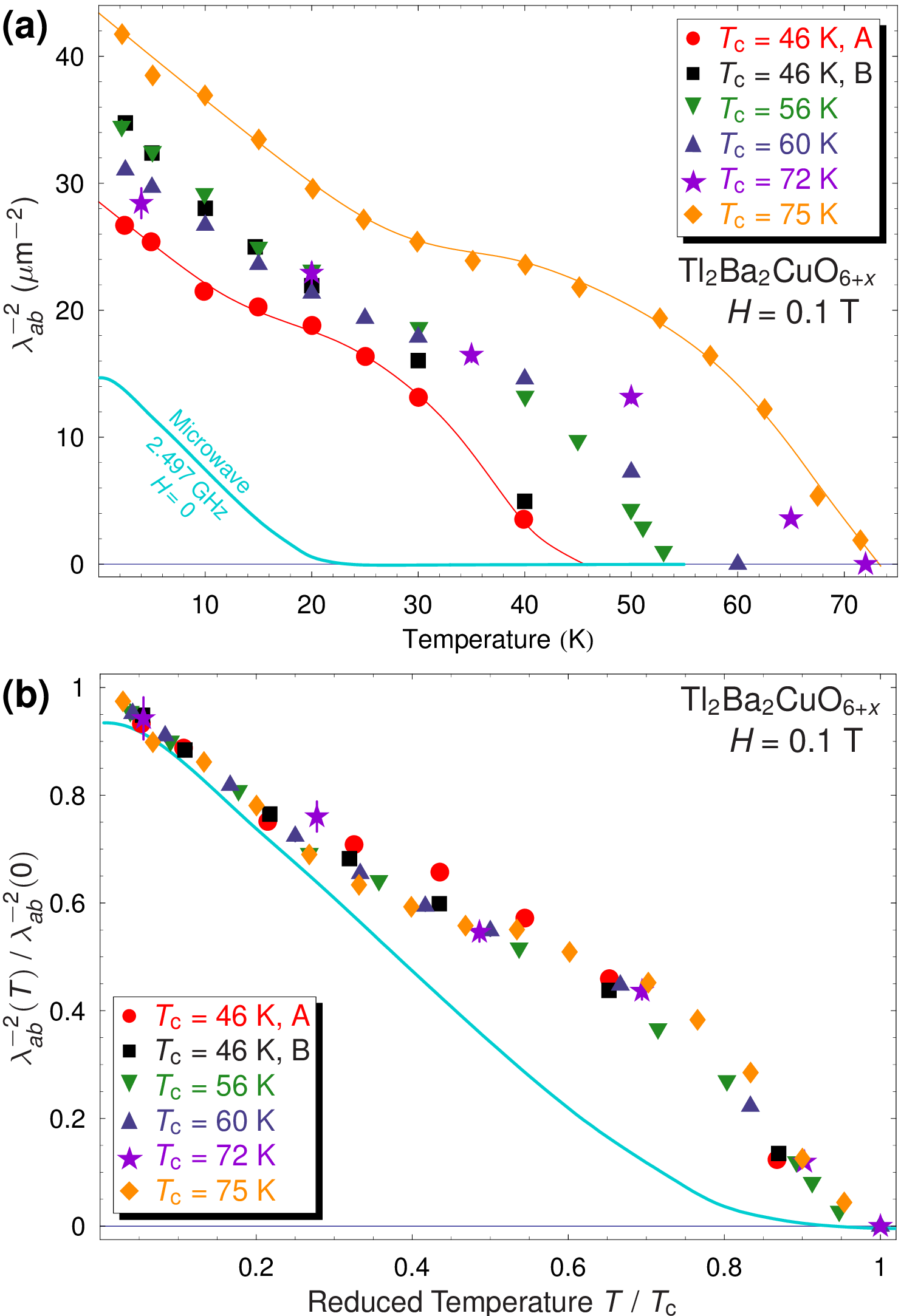} 
\caption{\label{fig:All-1kG-ILSQ_vs_T}
 (a) Temperature dependence of fitted $\lambda_{ab}^{-2}$ at
 $H=0.1$\,T for all Tl-2201 mosaics; A and B denote two different
 mosaics with the same \Tc.  Absolute microwave data (curve) at zero
 field on a $\Tc=25$\,K crystal at 2.497\,GHz \cite{Broun2014}, included
 for comparison, follow a qualitatively different form.  Curves are
 provided for two mosaics as a guide to the eye. (b) Normalized
 values $\lambda_{ab}^{-2}(T)/\lambda_{ab}^{-2}(0)$ {\it vs}.\ reduced
 temperature $T/\Tc$ for all Tl-2201 mosaics.  All dopings exhibit
 essentially the same temperature dependence, and differ from the
 microwave results (solid curve).
}
\end{figure}

\begin{table}[htb] 
\caption{\label{lambda-tab}Zero-temperature in-plane magnetic penetration
  depths in 0.1\,T for overdoped Tl-2201 mosaics having various \Tc s,
  from a linear extrapolation of $\lambda_{ab}^{-2}(T)$ at low
  temperatures, with estimated uncertainties in parentheses.  The
  variations in $\lambda_{ab}(0)$ are most likely dominated by the
  degree of order in the samples, rather than any systematic doping
  dependence, as discussed in the text.  Uncertainties in
  \Tc\ represent primarily the variation in \Tc\ among the crystals
  comprising the mosaic.
}
\begin{tabular}{|l||r|r|r|r|r|r|}\hline 
\Tc\ (K) & 46(1), A & 46(1), B & 56(1) & 60(1) & 72(1) & 75(1)\\ \hline 
$\lambda_{ab}(0)$ (nm) & 187(2) & 165(2) & 166(1) & 175(1) & 182(2) & 153(2)\\ \hline
\end{tabular}
\end{table}

Figure \ref{fig:All-1kG-ILSQ_vs_T} shows the vortex-state
$1/\lambda_{ab}^2(T)$, which is proportional to the superconducting
carrier density, for the six mosaics measured, and
Table~\ref{lambda-tab} reports the zero-temperature penetration depth
$\lambda_{ab}(0)$ from linear extrapolations of
$\lambda_{ab}^{-2}(T)$.  A highly unusual $T$-dependence, common to
all dopings, is immediately apparent.  The extent of this similarity
is more striking when $\lambda_{ab}^{-2}(T)$ is normalized to its
extrapolated $T=0$ value and plotted against reduced temperature
$T/\Tc$\ --- the {\sl relative} temperature dependence is {\sl
  identical}.  The most intriguing feature, exhibited in all six
mosaics, is upward curvature between $\frac{1}{3}\Tc$ and an
inflection point around 0.5\Tc.  This unusual temperature dependence
is robust and evident in any measure of the linewidth, but is absent
in zero-field (Meissner state) microwave surface resistance at higher
and lower dopings \cite{Ozcan2006,Broun2014}, the former included for
comparison.  The intrinsic $T$-dependence of the superconducting
carrier density (or $1/\lambda_{ab}^2$) in a single-gap $s$- or
$d$-wave superconductor exhibits downward curvature over the entire
temperature range 0--\Tc.


\section*{Discussion}

$\mu$SR reports of the cuprates' temperature-dependent in-plane
penetration depth typically exhibit the shape associated with a pure
$d$-wave order parameter~\cite{Sonier1994,Sonier1999}.  However, this
has not been the case in all data.  Upward curvature in the $\mu$SR
penetration depth can be recognized in relatively disordered overdoped
LSCO \cite{Khasanov2007}; at high dopings in cleaner
YBa$_2$Cu$_3$O$_{7-\delta}$ \cite{Sonier2007prb}, in lightly underdoped
YBa$_2$Cu$_4$O$_8$ \cite{Khasanov2008}; and in optimally and overdoped
Bi$_2$Sr$_2$CaCu$_2$O$_{8+\delta}$ \cite{Blasius1999}. This unusual
temperature dependence has appeared most clearly near and above
optimal doping \cite{Sonier1999,Sonier2007prb}, with some limited
evidence that it strengthens on overdoping~\cite{Blasius1999}.  It is
most evident at relatively low applied
fields~\cite{Harshman2004,Amin2000}.


With no phase transition expected at the fields and temperatures in
question, no such feature unambiguously visible in most published data
or any microwave penetration depth studies, and based on a small
number of data points in many of these cases, it has not been widely
accepted as a real effect.  Its now-confirmed appearance in a variety
of systems, and its particularly conspicuous appearance in Tl-2201,
implies that it is real and generic, at least to high doping ranges.
The contrast with zero-field microwave data argues against the few
interpretations floated thus far, instead pointing toward an origin in
an unexpected property of the cuprates' vortex state.  We first
briefly dispense with some alternative explanations before returning
to vortex physics.

First, multiband superconductors, those with more than one band
crossing the Fermi level, can exhibit unconventional temperature
dependence in the vortex state \cite{Zehetmayer2013}; a two-component
order parameter, {\it e.g.}\ $d+s$, with separate order parameters on
distinct Fermi surface sheets, has been advanced to explain the
LSCO \cite{Khasanov2007} and YBa$_2$Cu$_4$O$_8$ \cite{Khasanov2008}
results.  However, Tl-2201 has only one FS sheet, and the pure
$d$-wave symmetry and dissimilar microwave penetration depth at both
lower and higher dopings \cite{Broun1997,Broun2014} exclude such an
origin.  Second, the remarkable scaling seen in
Fig.~\ref{fig:All-1kG-ILSQ_vs_T} argues against an electronic phase
transition within the superconducting dome, such as an extension of
the pseudogap crossover temperature $T^*$.  Third, dilute paramagnetic
impurities would yield an additional broadening scaling with $H/T$,
contrary to the observed $T$ dependence.  Some type of magnetically
frozen state might account for the observed temperature dependence
(since $\mu^+$SR is a local probe, macroscopic phase separation would
not affect the superconducting component's lineshape). However, one
would not expect the onset of a competing magnetic phase to track
\Tc\ with doping \cite{Niedermayer98}. Furthermore, higher fields
should enhance any competing magnetic order \cite{Wu2011}, particularly
in the vortex cores \cite{Sonier2007}, but in YBCO they instead
suppress the exotic upward curvature in the temperature dependence of
the linewidth \cite{Harshman2004,Khasanov2007}.  Proximity-induced
chain superconductivity has been advanced as an explanation for the
inflection point in YBCO \cite{Sonier2007prb}, but this cannot explain
its appearance in chain-free Tl-2201.


Having excluded several alternative explanations, we return to physics
of the vortex phase, which would be absent in the Meissner-phase
microwave experiments and previous work on Tl-2201
powder \cite{Uemura1993}, and may thus offer a natural explanation.
First, the resistive upper critical field of Tl-2201 (actually the
irreversibility field \cite{Lundqvist1999,Zheng2000}) exhibits unusual
upward curvature \cite{Mackenzie1993} and stays very close to
$H_{c2}(T)$, far from the required temperature regime at the low
fields relevant here. A dimensionality crossover within the frozen
vortex state, as in the much more anisotropic
Bi$_2$Sr$_2$CaCu$_2$O$_8$ \cite{Bernhard1995,Blasius1999}, would
produce a symmetric lineshape, and the reduced field inhomogeneity
would narrow the linewidth at high temperatures, in contrast to our
results.  Trapping of vortices by preferred pinning sites at low
temperature has been advanced to explain the inflection in
YBCO \cite{Harshman2004}, but the linewidth changes in Tl-2201 would
require {\sl very} significant disorder and change the lineshape
significantly.  This is not seen in the FFT spectra; moreover,
allowing the temperature-independent $\sigma_0$ broadening to vary
produced no systematic trend with temperature. To further exclude
vortex disorder, it will be essential to quantify it independently
using, for instance, small-angle neutron scattering or scanning probe
techniques.

Another relevant feature of the vortex state is the symmetry of the
vortex lattice itself. An unusual square vortex lattice, as found in
optimally and overdoped LSCO \cite{Gilardi2002,Chang2012b} and in
YBCO \cite{White2011}, produces a broader but qualitatively similar
field distribution with a more pronounced low-field tail. However, the
vortex lattice in fully-oxygenated YBCO gradually transforms from
triangular to square as the field increases from 4 to 11\,T, while
strong upward curvature in the second moment of the TF-$\mu^+$SR
lineshape is strongest at much lower fields $\sim
0.1$\,T \cite{Khasanov2007b} and is completely suppressed by
4\,T \cite{Sonier1999}, excluding this interpretation, at least for
YBCO.  The $\mu$SR data for the $\Tc = 56$\,K Tl-2201 mosaic were fit
to a simple square vortex lattice model as a trial, but no crossover,
in the form of a systematic improvement in the quality of fit, was
evident.


We are thus drawn to the conclusion that the upward curvature must
arise from some fundamental property intrinsic to the $d$-wave
vortices themselves.  In this scenario, the difference between the
zero-field microwave measurements and these vortex-state $\mu$SR
results arises because the measurements probe different phases.  In
microwave measurements, only the surface contributes, while the bulk
is shielded.  In $\mu$SR, however, vortices also contribute, and these
can overlap and interact in certain field and temperature regimes.

The scaling with \Tc\ and absence in underdoped
samples point to an explanation in terms of the electronic structure
expected for a vortex in a $d$-wave superconductor. Theoretical
calculations \cite{Ichioka1999a,Ichioka1999b} show that such vortex
cores shrink with increasing magnetic field due to an enhanced
transfer of quasiparticles between neighbouring vortices. This effect
has been observed in YBCO by
$\mu$SR \cite{Sonier1999R,Sonier1999,Sonier2007prb}, with the core size
rapidly shrinking and saturating at fields above $H \sim 4$\,T.  At low
fields where the vortices are well separated, the quasiparticle
transfer is minimal and the vortices are expected to behave
essentially as if they were isolated. A vortex in a $d$-wave
superconductor is fourfold symmetric at low temperatures, with
extended low-energy quasiparticle states along the 45$^\circ$ (nodal)
directions in the CuO$_2$ plane. With increasing thermal population of
the higher energy states, the vortex core size grows, and calculations
show that above $\sim 0.5\Tc$ the fourfold-symmetric magnetic field
profile about the vortex core becomes nearly
cylindrical \cite{Ichioka1996b}.

As previously stressed \cite{Sonier2007}, $\lambda_{ab}$ as measured by
$\mu$SR may be regarded as the in-plane magnetic penetration depth
only in the $T \rightarrow 0$ and $H \rightarrow 0$ limit, but is
otherwise an effective length scale partially influenced by changes to
the field profile outside the vortex core by extended quasiparticle
states. Consequently, below $\sim 0.5\Tc$ and at low fields the
temperature dependence of $\lambda_{ab}$ is expected to be influenced
by the evolving quasiparticle states that extend far beyond the vortex
core. At higher temperatures, where the fourfold symmetry of the
vortex is essentially gone, the behavior of $\lambda_{ab}$ should
closely resemble that of the magnetic penetration depth, unless the
vortex lattice melts. Since the vortex core radius is directly
proportional to the gap magnitude \cite{Sonier2004review}, which in
turn is proportional to \Tc\ in overdoped cuprates, the scaling with
\Tc\ is naturally explained.

The inflection point near $0.5\Tc$ being less prominent or absent in
underdoped cuprates is likely due to the stabilization of competing
charge-density-wave (CDW) order localized in the vicinity of the
vortex cores. STM measurements on optimally- and slightly overdoped
Bi$_2$Sr$_2$CaCu$_2$O$_{8+\delta}$ \cite{Hoffman2002,Levy2005} and
nuclear magnetic resonance (NMR) measurements on underdoped
YBCO \cite{Wu2011,Wu2013} show static CDW order in the vortex core
region, where superconductivity is suppressed. While this induced
static CDW order is observed only at low temperatures in optimally and
slightly overdoped samples, in underdoped YBCO with $p \sim 0.12$, NMR
shows that static CDW order occurs over much of the temperature range
below \Tc. The occurrence of static CDW order in the vortex core
region constitutes a significant modification of the electronic
structure of the $d$-wave vortex \cite{Agterberg2015}, and consequently
the loss of the inflection point in the temperature dependence of
$\lambda_{ab}$ is not surprising.  The CDW competes with
superconductivity \cite{Chang2012,Ghiringhelli2012}, so it will
preferentially inhabit --- and likely gap out --- the regions of
momentum space in the nodal direction at the Fermi surface to
maximally avoid competition between the two orders.  As a result, the
extended quasiparticle core states along the nodal directions will
become bound.  This should lead to isotropic $s$-wave-like vortex
behaviour throughout much of the underdoped side of the phase diagram,
while the behaviour seen in the overdoped regime reflects the
intrinsic physics of $d$-wave vortices without such complications.
Supplementary Fig.~\ref{fig:shift} shows that the fit parameters,
based on an $s$-wave model, fail at the lowest temperatures (beginning
well below the inflection point).  The $s$-wave model's inability to
reproduce the observed field distribution provides evidence of its
failure to adequately describe the vortex phase, particularly at low
temperatures.

\begin{figure}[htb] 
\includegraphics[width=\columnwidth,clip]{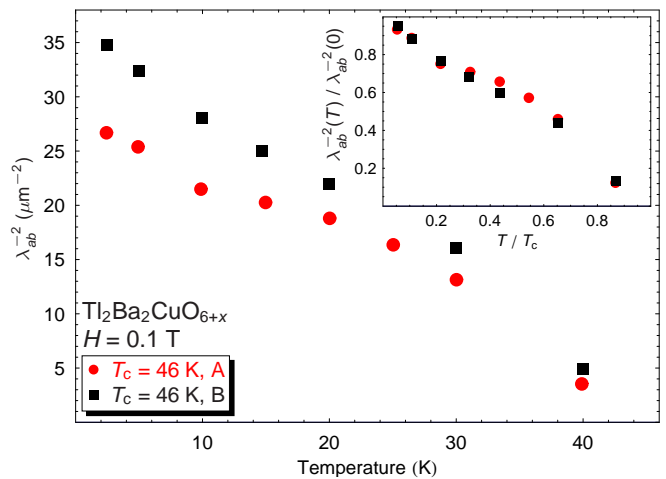} 
\caption{\label{fig:2005_vs_2008-Tc46_1kG-ILSQ_vs_T} Fitted values of
  $\lambda_{ab}^{-2}$ {\it vs}.\ $T$ for the earlier (``A'') and later
  (``B'') $\Tc = 46$\,K mosaics.  Inset: same data in normalized form.
} 
\end{figure} 

Finally, our use of mosaics with similar \Tc s has important
implications for techniques relying on $\mu$SR for values of the
zero-temperature penetration depth.  Mosaics grown and annealed under
very similar conditions, having the same or similar \Tc, exhibit quite
different absolute penetration depths, as shown in
Fig.~\ref{fig:2005_vs_2008-Tc46_1kG-ILSQ_vs_T} for $\Tc=46$\,K.  The
{\sl normalized} linewidths
$\lambda_{ab}^{-2}(T)/\lambda_{ab}^{-2}(0)$, however, are almost
identical.  Mosaics with \Tc s of 46\,K (``A'') and 72\,K were prepared
several years before the 46\,K (``B'') and 75\,K mosaics.  The crystal
growth was still being optimized when the early mosaics were
assembled, demonstrating that suppression of \Tc\ by disorder is {\sl
  not equivalent} to its suppression by carrier overdoping.  That the
zero-temperature $\lambda_{ab}^{-2}$ can apparently increase by $\sim
30$\% due to increasing crystalline perfection means that the
$\lambda_{ab}(0)$ values in Table~\ref{lambda-tab} should {\sl not} be
regarded as intrinsic.  A variety of other techniques rely upon
$\mu$SR to obtain absolute penetration depth values, but our work
indicates that in the overdoped regime, the values are only valid for
$T, H\rightarrow 0$, and even then are highly susceptible to disorder
and must be treated with caution.

The nature of superconductivity in the cuprates remains one of the
most important open questions in condensed matter physics, and the
overdoped regime proffers the promising prospect of understanding the
normal state from which high-temperature superconductivity emerges.
However, our $\mu^+$SR data indicate exotic physics survives to high
dopings.  A striking universal temperature dependence unambiguously
confirms an unusual upturn in the in-plane penetration depth, now seen
in at least four distinct material families, establishing that it is
new physics generic to the cuprates. Fundamental differences from
zero-field microwave measurements of nominally the same
quantity \cite{Broun2014} imply it is intrinsic to the $d$-wave
vortices themselves.  Aside from the substantial impacts of this
result on vortex physics and what light this may shed on cuprate
superconductivity, there are important ramifications for other
techniques.  $\mu$SR is uniquely suited to extracting the absolute
penetration depth from the vortex phase, and the values thus obtained
underpin results of other techniques which can't measure $\lambda$ in
absolute terms or at all.  While the $\mu$SR temperature evolution
should be robust \cite{Sonier2000}, the absolute values are
model-dependent.  Our data indicate that a more complex model is
required when $d$-wave vortices are present, which will necessitate
revisiting some previous results.  First, however, the details and
origin of the exotic temperature dependence must be conclusively
confirmed.  Sensitive scanning probe microscopies may provide insights
into the current and quasiparticle distribution in the vortex cores,
while confirmation that this is a vortex state phenomenon awaits
Meissner-state field distribution measurements, such as low-energy
$\mu$SR~\cite{Ofer2012} or $\beta$-NMR~\cite{Hossain2009}.  The
regimes in which these $d$-wave vortices manifest their unique
$d$-wave physics may provide crucial new insight on the cuprates'
order parameter and still-mysterious pairing.

\section*{Methods}

Single crystals of Tl-2201 were grown in gold-sealed alumina crucibles
by an encapsulated copper-rich self-flux method as described elsewhere
\cite{Peets2010}.  The oxygen content (which determines hole content
and Tc) was set by annealing under controlled oxygen partial pressures
and temperatures \cite{Opagiste1993}; two different annealing schemes
were employed depending on the desired oxygen
content~\cite{Peets2010}.  Crystals were assembled in mosaics on
substrates of aluminized mylar or GaAs to minimize the background
signal, with the crystallographic $c$-axis perpendicular to the
substrate (parallel to the applied field).  In this geometry, the
applied field is shielded by supercurrents running within the
$ab$-plane, thus the in-plane penetration depth $\lambda_{ab}$ governs
the field distribution.  The large number of small crystals to be
mutually aligned precluded measuring the superconducting transition of
every individual crystal, but care was taken to construct each mosaic
from a small number of annealing batches, each drawing crystals from
only one growth run, and \Tc\ was measured on a selection of crystals
sampled from each annealing run.  Examples of magnetization data on
three mosaics are included in Supplementary Fig.~\ref{fig:supp}.
Quoted uncertainties in $\Tc$ reflect transition widths of individual
crystals, as determined by DC magnetization in applied fields of
0.1--0.2\,mT, and the expected variation within the mosaic based on the
crystals sampled.  These fields were used because the lower critical
field $H_{c1}$ in this material is quite low.

Spin-polarized positive muons from the M15 muon channel at TRIUMF were
injected into the mosaics at an energy of 3-4\,MeV in one of several
different $\mu^+$SR spectrometers.  While a range of magnetic fields
were used, the lowest field for which the rotating reference frame
transformation worked reliably was 0.1\,T, and this field was used for
all data presented here.  In muon spin rotation/resonance/relaxation
($\mu$SR) \cite{Brewer1994,Sonier2000}, implanted muons settle into
specific preferred crystallographic sites, where their spins precess
around the local magnetic field $\vec B_{loc}$ with frequency $\omega
= \gamma B_{loc}$, where the muon gyromagnetic ratio $\gamma \approx
2\pi\times 135.54$\,MHz/T.  The precession is detected {\it via} a
decay positron (in the case of $\mu^+$) emitted preferentially along
the muon's spin direction.  The experimental $\beta^+$ decay asymmetry
reflects the precession of the ensemble of $\sim 10^7$ randomly
implanted muons and thereby the distribution of the local magnetic
field.  A Fourier transform of the time spectrum, which can be useful
for visualizing the field distribution characteristic of the vortex
state, shows a low-field cutoff corresponding to the midpoint of a
triad of near neighbor vortices, a Van Hove cusp from saddle points in
the vortex lattice, and a high-field tail and cutoff corresponding to
the maximum $B_{loc}$ in the vortex cores.  However, fitting is best
performed on the original time spectra: statistics decay with the muon
lifetime (2.197\,$\mu$s), and mixing high- with low-statistics data (as
in an FFT) is undesirable, but later times determine the frequency
resolution and thereby the reliability of fit parameters.  All
$\lambda_{ab}$ values were therefore extracted from fits in the time
domain to the lineshape described in Ref.~\onlinecite{Sonier2004} as
calculated numerically for a triangular vortex lattice with
$\lambda_{ab}$ and $\xi_{ab}$ as fitted parameters.  A test with a
square vortex lattice used the same approach.

\section*{Acknowledgements}

This work was supported by the Canadian Institute for Advanced
Research and the Natural Sciences and Engineering Research Council of
Canada.  The authors are indebted to Ch.\ Bernhard, D.M.\ Broun,
M.\ Franz, and K.\ Machida for stimulating discussions, D.\ Deepwell
and D.M.\ Broun for microwave data, M.D.\ Le and A.C.\ Shockley for
critical reading of the manuscript, and the TRIUMF CMMS staff for
their assistance.  A portion of this work was supported by
IBS-R009-G1.

\section*{Author Contributions}

This study was conceived, planned and supervised by JHB, DAB, WNH, and
RL. Crystals were grown, annealed and characterized by DCP and RL,
then aligned into mosaics by SLS, JHB and RL.  $\mu^+$SR experiments
were performed by SLS and JHB, and JHB was pestered by DCP and WAM
into analysing the data.  The data were interpreted and the manuscript
written by WAM, DCP, JHB, and JES.

\section*{Additional Information}

The authors declare no competing financial interests.

\section*{References}

\bibliography{short_paper.bbl}

\appendix
\input{supp.tex}

\end{document}

%% file: supp.tex
\renewcommand{\thefigure}{S\arabic{figure}}
\setcounter{figure}{0}
\section{Supplementary Information}

\begin{figure}[htb]
\includegraphics[width=\columnwidth]{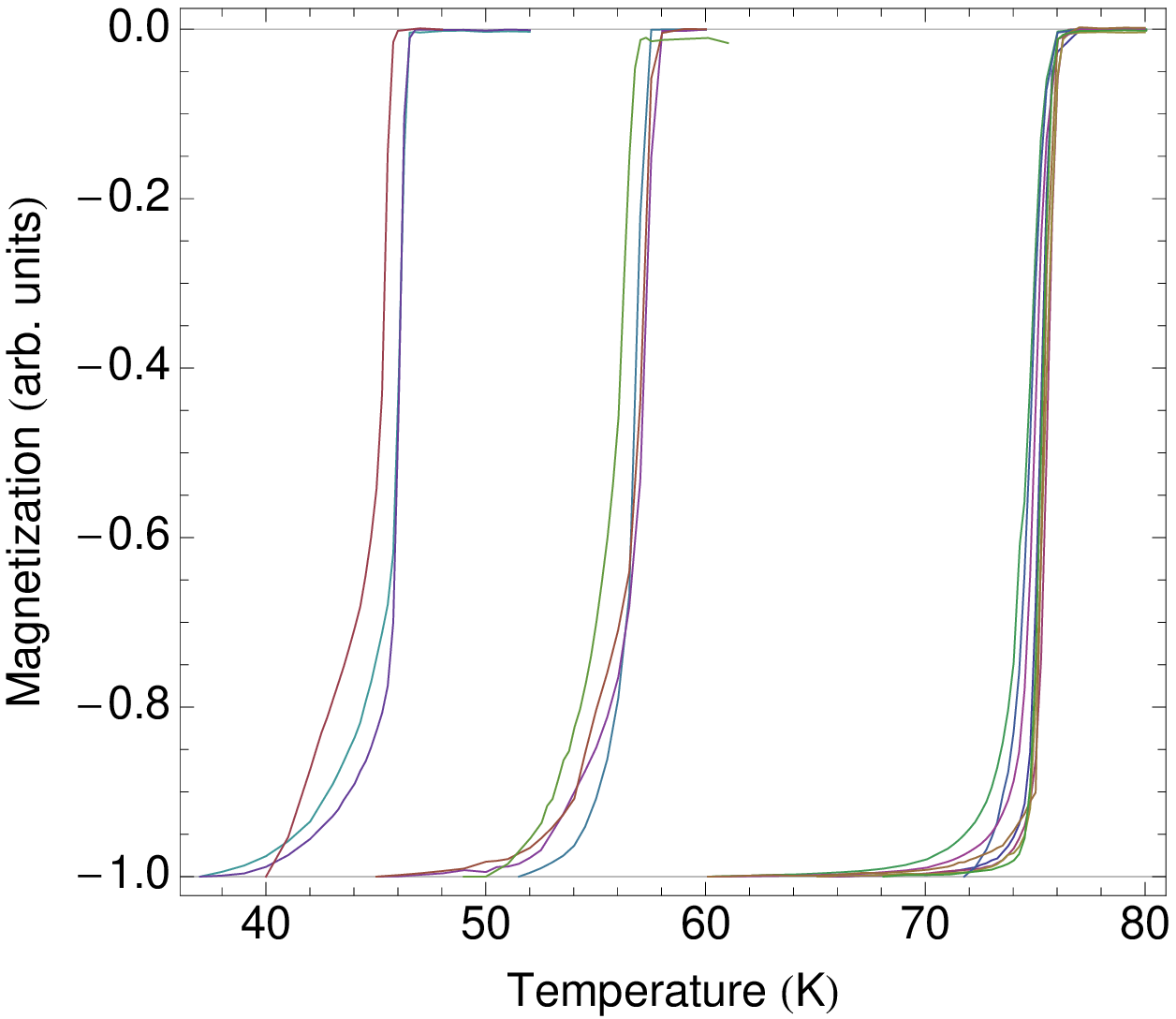}
\caption{\label{fig:supp}Example field-cooled magnetization data for
  three of the mosaics measured in this study.}
\end{figure}

The field-cooled magnetization data collected for three mosaics ---
46\,K (B), 56\,K, and 75\,K --- are shown in Fig.~\ref{fig:supp}.
Data were taken on cooling in applied fields of 0.1--0.2\,mT to
minimize broadening due to the lower critical field.  To aid
comparison, the data have been normalized.  All data were collected
with the field applied along the $c$-axis.

\begin{figure}[htb]
\includegraphics[width=\columnwidth]{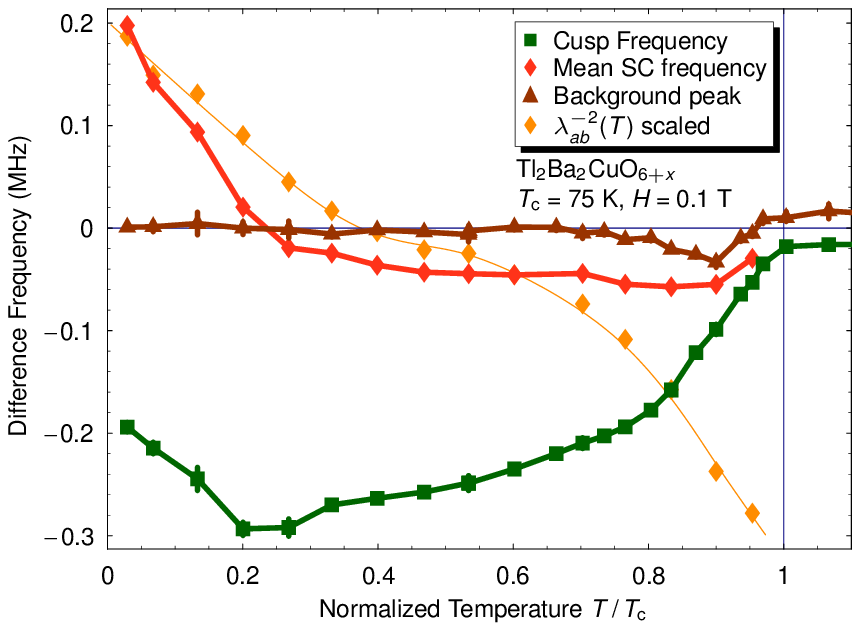}
\caption{\label{fig:shift} Anomalous temperature-dependence of fit
  parameters at $H=0.1$\,T for the $\Tc=75$\,K mosaic.  The frequency
  of the background peak, superconducting cusp, and the mean frequency
  (first moment) of the fitted vortex lattice lineshape are shown,
  with the frequency corresponding to the applied field subtracted.
  Rescaled $\lambda_{ab}^{-2}(T)$ values from
  Fig.~\ref{fig:All-1kG-ILSQ_vs_T} are included for reference, again
  with a curve to serve as a guide to the eye.}
\end{figure}

Figure~\ref{fig:shift} shows anomalous behaviour in the extracted
vortex lattice lineshape, using data collected on the $\Tc=75$\,K
mosaic at $H=0.1$\,T as a representative example.  Because the field
is applied by a driven superconducting magnet and can drift by up to
0.5\% over the course of a full temperature sweep, frequency values
here are corrected by subtracting the frequencies corresponding to the
actual applied fields as measured by a Hall sensor.  The background
peak only drifts significantly near \Tc, where it becomes difficult to
distinguish from the superconducting signal.  The mean frequency
(first moment) of the fitted superconducting lineshape drifts higher
at low temperature, while the cusp corresponding to saddle points
between vortices departs from the background on cooling as expected
before coming back toward it.  In both cases, the departure from
expected behaviour occurs at temperatures well below the inflection
point in the extracted $\lambda_{ab}^{-2}(T)$.  Other fit parameters
exhibited no systematic temperature dependence.  The anomaly reflects
a subtle change in shape of the superconducting field distribution,
indicating the breakdown at low temperatures of the $s$-wave vortex
model used, and may provide guidance as to the temperature regime in
which $d$-wave vortex physics must be taken into account.

Raw data for all $\mu$SR measurements performed at TRIUMF are freely
available at \href{http://musr.physics.ubc.ca/mud}{\tt
  http://musr.physics.ubc.ca/mud}.  Data used in this study may be
found by searching for Experiment 958.  The current version of
{\sffamily LSHfit}, used to analyse the data, is available from Jeff
Sonier on request.